\documentclass[aps,prb,twocolumn,showpacs,twoside,floatfix,superscriptaddress]{revtex4}

\usepackage{graphicx}
\usepackage{graphics}
\usepackage{amsmath}
\usepackage{amssymb}
\usepackage{mathrsfs}
\usepackage{epsf}

\begin{document}

\title{Transport and bistable kinetics of a Brownian particle in a nonequilibrium environment}

\author{Jyotipratim Ray Chaudhuri}
\email{jprc_8@yahoo.com}
\affiliation{Department of Physics, Katwa College, Katwa, Burdwan 713130, India}

\author{Suman Kumar Banik}
\affiliation{Department of Biological Sciences, Virginia
Polytechnic Institute and State University, Blacksburg, VA
24061-0406, USA}

\author{Sudip Chattopadhyay}
\email{sudip_chattopadhyay@rediffmail.com}
\affiliation{Department of Chemistry, Bengal Engineering and Science University, Shibpur, Howrah 711103, India}

\author{Pinaki Chaudhury}
\affiliation{Department of Chemistry, University of Calcutta,
Kolkata 700009, India}

\date{\today}

\begin{abstract}
A system reservoir model, where the associated reservoir is
modulated by an external colored random force, is proposed to
study the transport of an overdamped Brownian particle in a
periodic potential. We then derive the analytical expression for
the average velocity, mobility, and diffusion rate. The bistable
kinetics and escape rate from a metastable state in the overdamped
region are studied consequently. By numerical simulation we then
demonstrate that our analytical escape rate is in good agreement
with that of numerical result.
\end{abstract}


\maketitle


\section{Introduction}
As an immediate consequence of stochastic dynamics, it is observed
that thermal diffusion in a periodic potential plays a prominent
role in various cases such as Josephson's junction \cite{refja},
diffusion in crystal surface \cite{refjb}, noise limit cycle
oscillators \cite{refjc} etc. There has been a renewed interest in
recent times in the study of transport properties of Brownian
particles moving in a periodic potential \cite{refjd} with special
emphasis on coherent transport and giant diffusion \cite{refje}.
These studies have been motivated in part by an attempt to
understand the mechanism of movement of protein motors in
biological systems \cite{refjf}. Several physical models have been
proposed to understand the transport phenomena in such systems
such as vibrational ratchet \cite{refjg}, rocking ratchet
\cite{refjh}, diffusion ratchet \cite{refji,refjj}, correlation
ratchet \cite{refjk}, etc. Such ratchet models have a wide range
of application in biology and nanoscopic systems \cite{refjl},
because of their extraordinary success in exploring experimental
observations on biochemical molecular motors, active in muscle
contraction \cite{refjm}, observation of directed transport in
photovoltaic and photoreflective materials \cite{refjn}, etc. In
all the above models, the potential is taken to be asymmetric in
space. One can also obtain a unidirectional current in the
presence of spatially symmetric potential. For such
non-equilibrium systems, one requires time asymmetric random force
\cite{refjo} or space dependent diffusion
\cite{refjp,refjq,refjr}. In passing, we want to mention the fact
that to explain the role of Levy stable noise in transport
phenomena in presence of bistable, metastable, and periodic
potentials with broken symmetry, several elegant approaches have
been suggested recently \cite{pre}.

Traditionally, Langevin equation describing the dynamics of a
Brownian particle coupled to a thermal reservoir is a tool for
modeling several aspects of nonequilibrium phenomena \cite{lax}.
In addition to the Langevin dynamics, one often takes into account
the probabilistic aspect of the random dynamics through the usage
of Fokker-Planck equation \cite{risken}. Both of the approaches
utilize the relation between the random fluctuations imposed by
the reservoir into the Brownian particle and the relaxation of the
imposed energy back to the reservoir, through the
fluctuation-dissipation relation (FDR) \cite{kubo}. FDR takes into
account the balance between the energy input (to the system from
the reservoir) and output (from the system into the reservoir)
through the detailed balance mechanism. Typical signature of such
a system is the attainment of equilibrium in the asymptotic limit.

An additional external random driving applied to the Brownian
particle can break this balance mechanism and make the composite
system open \cite{lw}, the direct consequence of which is the loss
of FDR. In addition to that, the system hardly reaches the
equilibrium state in the long time limit but rather attains a
stationary steady state \cite{lw}. However, a driving of the
reservoir by an external random field \cite{bravo} creates a
thermodynamic consistency condition analogous to FDR
\cite{jrcpre1} that leads to the study of several interesting
phenomena \cite{jrcpre1,jrcpre2,jrcpre3,jrcjpa,jrcjcp} in chemical
physics. The net effect of the reservoir driving by an external
random force is the creation of an effective temperature in
addition to the thermal energy $k_BT$ exerted by the reservoir on
the system of interest. As shown recently, this effective
temperature can enhance the reaction rate in condensed media
\cite{jrcpre1,jrcpre2,jrcpre3} as well as generate directed motion
in a periodic system \cite{jrcjpa,jrcjcp}.

In the present paper, we consider a system-reservoir model where
the bath is modulated by an external noise. However, when the
reservoir is modulated by an external noise  it is likely that it
induces fluctuations in the polarization of the reservoir due to
the external noise from a microscopic point of view and one may
expect that the non-equilibrium situation created by modulating
the bath makes its presence felt in the transport property and
also in the kinetics of the Brownian particle. A number of
different situations depicting the modulation of the bath may be
physically relevant. For an example, we may consider the case of a
Brownian particle when the response of the solvent is
time-dependent, as in a liquid crystal, or in the
reaction-diffusion mechanism in super-critical lattice, or the
growth in living polymerization \cite{refjrr}.

To observe the effects of external stochastic modulation one can
carry out the experiment in a photochemically active solvent (the
heat bath) where the solvent is under the influence of external
monochromatic light with fluctuating intensity which is absorbed
solely by the solvent molecules. As a result of this, the
modulated solvent heats up due to the conversion of light energy
into heat energy by radiation less relaxation process, and an
effective temperature-like quantity develops due to constant input
of energy. Since the fluctuations in light intensity result in the
polarization of the solvent molecules the effective reaction field
around the reactant gets modified \cite{refjss}. Our theoretical
model can be tested experimentally to study the directional motion
and mean first passage time of artificial chemical rotors in
photovoltaic solvent \cite{refjs}.

There are some precedents for our model that are worth mentioning.
Bravo {\it et al.} \cite{bravo} dealt with a related problem: a
classical system in a heat bath with an additive external noise.
In the quantum case, Faid and Fox \cite{refjtt} proposed a
stochastic coupling between the system and a heat bath as a
phenomenological mechanism of relaxation for the bath. Ma\~{n}as
{\it et al.} \cite{refju} considered the problem of a system
coupled to an ensemble of independent harmonic oscillators as a
reservoir. We have also considered the dynamics of a metastable
state non-linearly coupled to a heat bath driven by an external
noise to study the escape rate from a metastable state
\cite{jrcpre3}. In the present work, we address the so-called
ratchet problem and bistable kinetics of a Brownian particle for a
thermodynamically open system where the associated bath is
modulated by a colored noise and we explore the dependence of
various parameters of the external noise on the transport
phenomena and bistable kinetics.

To get insight of situations encountered in the
growing number of nano and microscale experiments, our model can
be used as a potential tool. For instance, several DNA nanomotors
have been recently suggested \cite{epl,epl8}. These machines are
relatively slow and do not perform continuous rotation. Very
recently, a rotary DNA nanomachine that shows a continuous
rotation has been proposed \cite{epl}. This motor consists of a
DNA ring whose elastic features are tuned such that it can be
externally driven by a periodic temperature change. Our model
proposed in this paper can be used as a theoretical avenue to
examine the periodic temperature change via a physically motivated
microscopic Hamiltonian picture. Optical tweezers \cite{t1}, which
is capable of manipulating nanometer and micrometer sized
dielectric particles by exerting extremely small forces via a
highly focused laser beam, is often used to manipulate and study
single molecules by interacting with a bead that has been attached
to that molecule. DNA and the proteins and enzymes that interact
with it are commonly studied in this way. At this juncture, we
want to mention the fact that our development present in this
article can be used as a theoretical model to understand the
phenomena of heating of the liquid surrounding a bead in an
optical tweezers setup \cite{t13}.

The organization of the paper is as follows: In Sec. \ref{model},
starting from a microscopic Hamiltonian picture of a system
linearly coupled with a harmonic reservoir which is modulated by a
noise with arbitrary decaying memory kernel, we have derived the
Langevin equation with an effective noise $\eta(t)$ and then
explored its statistical property. Employing the functional
calculus method \cite{lw,refjw,refjy}, we then obtain the
Fokker-Planck-Smoluchowski equation in Sec. \ref{fokker},
corresponding to the Langevin equation valid in the over damped
limit and for rapid fluctuations whose correlation function
vanishes rapidly. In Sec. \ref{application}, we have calculated
the steady current in a ratchet potential and derived the
expression for diffusion rate and mobility. As another application
of our development, we study the bistable kinetics to obtain the
stationary probability density function (PDF) and the barrier
crossing rate. The summarizing remarks are presented in Sec.
\ref{conclusion} preceded by a numerical application in Sec.
\ref{secnum}.


\section{The Model: Heat bath modulated by external
noise}\label{model}

We consider a classical particle of unit mass coupled to a heat
bath consisting of a set of $N$-numbers of mass weighted harmonic
oscillators with frequency $\{\omega_j\}$. The heat-bath is
externally driven by a Gaussian random force $\epsilon(t)$ with
an arbitrary decaying correlation function. The total Hamiltonian
of such a composite system can be written as \cite{zwanzig,jrcpre1}
\begin{eqnarray}\label{eq1}
H = \frac{v^2}{2} + V({x}) + \sum_{j=1}^N \left [
\frac{{v}_j^2}{2}
+\frac{1}{2} \omega_j^2 ( x_j - c_j x)^2 \right]
+H_{\rm int}.
\end{eqnarray}

\noindent In the above equation $x$ and $v$ are the coordinate and
the velocity of the system particle, respectively, and $V(x)$ is
the potential energy of the system. $\{x_j,v_j\}$ are the
variables for the $j$th oscillator with characteristic frequency
$\omega_j$. The system particle is coupled to the bath oscillator
linearly through the general coupling terms $c_j\omega_j x$ where
$c_j$ is the coupling strength for the system-bath interaction.
The interaction between the heat bath and the external noise is
represented by the term $H_{\rm int}$ which we take as
\cite{jrcpre1,landau}
\begin{eqnarray}\label{eq2}
H_{\rm int} = \sum_{j=1}^N  \kappa_j x_j \epsilon(t),
\end{eqnarray}

\noindent where $\kappa_j$ denotes the coupling strength of
interaction and $\epsilon(t)$ is an external noise which is
assumed to be stationary, Gaussian with zero mean and
arbitrary decaying correlation function, the statistical property
of which is given by
\begin{eqnarray}\label{eq3}
\langle \epsilon(t) \rangle_e = 0,
\langle \epsilon(t) \epsilon(t^{\prime}) \rangle_e = 2 D_e \Psi(t-t^{\prime}),
\end{eqnarray}

\noindent where $D_e$ is the external noise strength and $\Psi(t)$
is the external noise memory kernel which is assumed to be a
decaying function of its argument and $\langle \cdots \rangle_e$
implies averaging over each realization of $\epsilon(t)$.
Eliminating the bath degrees of freedom in the usual way
\cite{zwanzig,jrcpre1}, we get the Langevin equation for the
system particle
\begin{eqnarray}\label{eq4}
\dot{x} &=& {v}, \nonumber \\
\dot{v} &=& - V^{\prime}({x}) - \int_0^t dt^{\prime}
\gamma(t-t^{\prime}) v(t^{\prime}) + f(t) + \pi(t),
\end{eqnarray}

\noindent where the memory kernel $\gamma(t)$ and the Langevin
force term $f(t)$ are given, respectively, by
\begin{eqnarray}\label{eq5}
\gamma(t) &=& \sum_{j=1}^N c_j^2 \omega_j^2 \cos(\omega_j t), \\
\label{eq6}
f(t) &=&  \sum_{j=1}^N c_j\omega_j^2 \left \{ [x_j(0) -c_j x(0)]
\cos(\omega_j t) \right. \nonumber \\
&& \left. + \frac{v_j (0)}{\omega_j} \sin (\omega_j t) \right \} .
\end{eqnarray}

\noindent In Eq.(\ref{eq4}),
$\pi(t)$ is the fluctuating force generated due to the external
stochastic forcing of the bath by $\epsilon(t)$ and is given by
\begin{equation}\label{eq7}
\pi(t) = - \int_0^t dt^{\prime} \varphi(t-t^{\prime})
\epsilon(t^{\prime}),
\end{equation}

\noindent with
\begin{equation}
\varphi(t) = \sum_{j=1}^N c_j\omega_j \kappa_j \sin(\omega_j t).
\end{equation}

The form of Eq.(\ref{eq4}) reveals that the system is driven by
two fluctuating forces, $f(t)$ and $\pi(t)$. $\pi(t)$ is a dressed
noise originating due to the bath modulation by external noise
$\epsilon(t)$, and $f(t)$ is the thermal noise due to system-bath
coupling. To define the statistical properties of $f(t)$, we
assume that the initial distribution is such that the bath is
equilibrated at $t=0$ in the presence of the system but in the
absence of the external noise $\epsilon(t)$ such that
\begin{eqnarray}\label{eq8}
\langle f(t) \rangle = 0, \langle f(t) f(t^{\prime})
\rangle = k_B T \gamma(t-t^{\prime}),
\end{eqnarray}

\noindent where $k_B$ is the Boltzmann constant and $T$ is the
equilibrium temperature, $\langle \cdots \rangle$ implies the
usual average over the initial distribution which is assumed to
be a canonical distribution of Gaussian form \cite{zwanzig,jrcpre1}
\begin{equation*}
\mathbb{P} = \mathcal{N} \exp \left \{- \frac{v_j^2(0) +
\omega_j^2 (x_j(0) -c_j x(0))^2}{2k_BT} \right \},
\end{equation*}

\noindent where $\mathcal{N}$ is the normalization constant. Now
at $t=0_+$, the external noise agency is switched on to modulate
the bath. Here, we define an effective Gaussian noise $\eta(t)=
f(t)+\pi(t)$ the statistical property of which can be described by
\begin{eqnarray}\label{eq9}
\langle \langle \eta(t) \rangle \rangle &=& 0, \nonumber \\
\langle \langle \eta(t) \eta(t^{\prime}) \rangle \rangle &=& k_BT \gamma
(t-t^{\prime}) + 2 D_e \int_0^t dt^{\prime\prime}
\int_0^{t^{\prime}}dt^{\prime\prime\prime} \nonumber \\
&& \times \varphi(t-t^{\prime\prime})
\varphi(t^{\prime}-t^{\prime\prime\prime})
\Psi(t^{\prime\prime}-t^{\prime\prime\prime}), \\
& =& G(t-t^{\prime}) \text{ (say)}. \nonumber
\end{eqnarray}

\noindent In Eq.(\ref{eq9}), $\langle \langle \cdots \rangle
\rangle$ means that we have taken two averages independently,
average over initial distribution of bath variables and average
over each realization of $\epsilon(t)$. While deriving
Eq.(\ref{eq9}), we have made the assumption $\langle \langle
\eta(t) \eta(t^{\prime})\rangle \rangle = G(t-t^{\prime})$, which
can not be proved unless the structure of $\varphi(t)$ is
explicitly given. As we shall see later it is a valid assumption
for a particular choice of coupling coefficients and for external
stationary noise processes. It should be realized that
Eq.(\ref{eq9}) is not a fluctuation-dissipation relation (FDR) due
to the appearance of the external noise intensity, rather it
serves as a thermodynamic consistency relation.

To obtain a finite result in the continuum limit, i.e., for $N
\rightarrow \infty$, the coupling function $c_i= c(\omega)$ and
$\kappa_i=\kappa(\omega)$ are chosen as $c(\omega) = c_0/\omega
\sqrt {\tau_c}$ and $\kappa(\omega) = \kappa_0 \omega \sqrt
{\tau_c}$. Consequently, $\gamma(t)$ and $\varphi(t)$ reduce to
\begin{eqnarray}\label{eq10}
\gamma(t) = \frac{c_0^2}{\tau_c} \int d \omega \rho(\omega) \cos
\omega t,
\end{eqnarray}

\noindent and
\begin{eqnarray}\label{eq11}
\varphi(t) = c_0 \kappa_0 \int d \omega \rho(\omega) \omega \sin
\omega t,
\end{eqnarray}

\noindent where $c_0$ and $\kappa_0$ are constants and $\tau_c$ is
the correlation time of the heat bath. For $\tau_c \rightarrow 0$ we
obtain a $\delta$-correlated noise process. $1/\tau_c$ may
be characterized as the cut off frequency of the bath oscillators.
$\rho(\omega)$ is the density of modes of the heat bath which is
assumed to be Lorentzian
\begin{eqnarray}\label{eq12}
\rho(\omega) = \frac{2 \tau_c}{\pi (1+\omega^2\tau_c^2)}.
\end{eqnarray}

\noindent The above assumption resembles broadly the behavior of
the hydrodynamical modes \cite{refjz,refjzz} in a microscopic
system and is frequently used by the chemical physics community
\cite{refjz}. With these forms of $\rho(\omega)$, $c(\omega)$ and
$\kappa(\omega)$, we have the expression for $\varphi(t)$ and
$\gamma(t)$, respectively, as
\begin{eqnarray}\label{eq13}
\varphi(t) = \frac{c_0\kappa_0}{\tau_c} \exp(-|t|/\tau_c),
\nonumber \gamma(t) = \frac{c_0^2}{\tau_c} \exp(-|t|/\tau_c).
\end{eqnarray}

\noindent Though Eq.(\ref{eq9}) is not a FDR, Eq.(\ref{eq7})
resembles the familiar linear relation between the polarization
and the external field. Here, $\pi(t)$ and $\epsilon(t)$ play the
role of former and latter, respectively. Thus $\varphi(t)$ can be
interpreted as a response function of the reservoir due to
external noise $\epsilon(t)$. It is also clear from the structure
of $\varphi(t)$ and $\gamma(t)$ that
\begin{eqnarray}\label{eq14}
\frac{d\gamma(t)}{dt} = - \frac{c_0}{\kappa_0} \frac{1}{\tau_c}
\varphi(t).
\end{eqnarray}

\noindent The above relation is independent of $\cal {D}(\omega)$ and
represents how the dissipative kernel $\gamma(t)$ depends on the
response function of the medium due to the external noise
$\epsilon(t)$. Such an equation for the open system can be
anticipated in view of the fact that both the dissipation and
response functions crucially depend on the properties of the
reservoir. If we assume that $\epsilon$ is a $\delta$-correlated
noise, i.e., $\langle \epsilon(t) \epsilon(t^{\prime}) \rangle_e =
2 D_e \delta(t-t^{\prime})$ then the correlation function of
$\pi(t)$ is given  by
\begin{eqnarray}\label{eq15}
\langle \pi(t) \pi(t^{\prime}) \rangle_e = \frac{D_e c_0^2
\kappa_0^2}{\tau_c} \exp(-|(t-t^{\prime})|/\tau_c),
\end{eqnarray}

\noindent where we have neglected the transient terms
$(t,t^{\prime}>\tau_c)$. This equation shows how the heat bath
dresses the external noise. Though the external noise is a
$\delta$-correlated one, the system encounters it as an
exponentially correlated noise with the same correlation time of
the internal noise but with a strength dependent on the coupling term
$\kappa_0$ and the external noise strength $D_e$. On the other hand,
if the external noise follows Ornstein-Uhlenbeck process
\begin{eqnarray*}
\langle \epsilon(t) \epsilon(t^{\prime}) \rangle_e =
\frac{D_e}{\tau_c} \exp(-|(t-t^{\prime})|/\tau_e),
\end{eqnarray*}

\noindent the correlation  function of $\pi(t)$ is found to be
\begin{eqnarray}\label{eq16}
\langle \pi(t) \pi(t^{\prime}) \rangle_e &=& \frac{D_e c_0^2
\kappa_0^2}{(\tau_e/\tau_c)^2 -1}
\left(\frac{\tau_e}{\tau_c} \right) \left \{ \frac{1}{\tau_c}
\exp(-|(t-t^{\prime})|/\tau_e) \right. \nonumber \\
&& \left. - \frac{1}{\tau_e}
\exp(-|(t-t^{\prime})|/\tau_c) \right \},
\end{eqnarray}

\noindent where we have again neglected the transient terms. The
dressed external noise $\pi(t)$ now has a more complicated
structure of correlation function with two correlation times
$\tau_c$ and $\tau_e$. If the external noise correlation time is
much larger than the internal noise correlation time (
$\tau_e>>\tau_c$), which is more realistic, the dressed noise is
dominated by the external noise and we have
\begin{eqnarray}\label{eq17}
\langle \pi(t)\pi(t^{\prime}) \rangle_e = \frac{D_e c_0^2
\kappa_0^2}{\tau_e} \exp(-|(t-t^{\prime})|/\tau_e).
\end{eqnarray}

\noindent On the other hand, when the external noise correlation time is
smaller than the internal one, we recover Eq.(\ref{eq15}). In
what follows, we shall focus on the situation when $\tau_e>>
\tau_c$. Thus, in terms of the effective noise $\eta(t)$, the
Langevin Eq.(\ref{eq4}) can be written as
\begin{eqnarray}\label{eq18}
\dot {x} &=& v, \nonumber \\
\dot{v} &=& -V^{\prime}(x) - \int_0^t
dt^{\prime} \gamma (t-t^{\prime}) v(t^{\prime}) + \eta(t),
\end{eqnarray}

\noindent which reduces to
\begin{eqnarray}\label{eq19}
\dot {x} = v,
\dot{v} = -V^{\prime}(x) - \gamma v(t) + \eta(t),
\end{eqnarray}

\noindent where we have assumed that the internal noise $f(t)$ is
$\delta$-correlated and the internal dissipation is Markovian  so
that
\begin{eqnarray*}
\gamma(t) &=& 2c_0^2 \delta (t-t^{\prime}) = 2 \gamma \delta
(t-t^{\prime}) \\
\langle f(t) \rangle &=& 0, \\
\langle f(t)  f(t^{\prime}) \rangle &=& 2c_0^2 k_BT
\delta(t-t^{\prime}) = 2 \gamma k_BT \delta (t-t^{\prime}),
\end{eqnarray*}

\noindent with $\gamma=c_0^2$ [see Eqs.(\ref{eq5}), (\ref{eq6})
and (\ref{eq7})]. The effective noise $\eta(t)$ thus has
statistical properties [see Eq.(\ref{eq8})]
\begin{eqnarray}\label{eq20}
\langle \langle \eta(t) \rangle \rangle &=& 0, \nonumber \\
\langle \langle \eta(t) \eta(t^{\prime})\rangle \rangle
&=& \frac{D_R}{\tau_R} \exp(-|(t-t^{\prime})|/\tau_R),
\end{eqnarray}

\noindent where
\begin{eqnarray}\label{eq21}
D_R = \gamma(k_BT + D_e \kappa_0^2) \text{ and }
\tau_R = \frac{D_e }{D_R} \gamma \kappa_0^2 \tau_e,
\end{eqnarray}
\noindent with $D_R$ and $\tau_R$ being the strength and
correlation time of the effective noise $\eta(t)$, respectively.
Although the reservoir is driven by the colored noise $\epsilon
(t)$ with noise strength $D_e$ and correlation time $\tau_e$, the
dynamics of the system of interest is effectively governed by the
scaled colored noise $\eta (t)$, with noise strength $D_R$ and
correlation time $\tau_R$. In what follows we will describe the
effect of external noise in terms of the effective parameters
$D_R$ and $\tau_R$ in the rest of our analysis.


\section{The Fokker-Planck description in the overdamped limit}\label{fokker}

In the over damped limit, Eq. (\ref{eq19}) reads
\begin{eqnarray}\label{eq22}
\dot{x}(t) &=& -\frac{1}{\gamma} V^{\prime}(x)+ \frac{1}{\gamma} \eta(t), \\
&=& W(x) + \xi(t), \label{eq23}
\end{eqnarray}

\noindent where we have defined $W(x) = -(1/\gamma) V^{\prime}(x)$
and $\xi(t)= (1/\gamma) \eta(t)$. Clearly, $\xi(t)$ is the scaled
noise and as $\eta(t)$ is Gaussian [since $f(t)$ and $\epsilon(t)$
are assumed to Gaussian] $\xi (t)$ is also Gaussian. The Gaussian
nature of $\epsilon(t)$ is expressed by the probability
distribution function
\begin{eqnarray}\label{eq24}
P [\xi(t)] = N \exp \left [-\frac{1}{2}  \int ds \int ds^{\prime}
\xi(s) \xi(s^{\prime}) \beta (s-s^{\prime}) \right],
\end{eqnarray}

\noindent where $\beta$ is the inverse of the correlation function
of $\xi(t)$, and $N$ is the normalization constant expressed by a
path integral over $\xi(t)$,
\begin{eqnarray}\label{eq25}
\frac{1}{N} = \int {\mathcal{D}} \xi \exp \left [-\frac{1}{2} \int
ds \int ds^{\prime} f(s) f(s^{\prime})\beta (s-s^{\prime})
\right].
\nonumber \\
\end{eqnarray}

\noindent Now, let $\langle \xi(t)\rangle =0$ and $\langle \xi(t)
\xi(t^{\prime})\rangle =c (t-t^{\prime})$. Then from
Eq.(\ref{eq25}) we get
\begin{eqnarray}
\frac{\delta N}{\delta \xi(t)} &=& - N^2 \int {\mathcal{D}} \xi
\left[-\frac{1}{2} \int ds^{\prime} \xi(s^{\prime}) \beta
(t-s^{\prime}) \right. \nonumber \\
&& \left. + \int ds f(s) \beta (s-t) \right] \nonumber \\
&& \times \exp \left [-\frac{1}{2} \int ds \int ds^{\prime} \xi(s)
\xi(s^{\prime}) \beta (s-s^{\prime}) \right], \nonumber \\
&=& N\int ds \beta(t-s) \langle \xi(s) \rangle = 0.  \label{eq26}
\end{eqnarray}

\noindent Therefore, it follows that
\begin{eqnarray}\label{eq27}
\frac{\delta P [\xi(t)]}{\delta \xi(t)} = - \left[ \int ds
\beta(t-s) \xi(s) \right] P[\xi].
\end{eqnarray}

\noindent Consequently,
\begin{eqnarray}\label{eq28}
\frac{\delta^2 P [\xi(t)]}{\delta \xi(t)\delta \xi(t^{\prime})} &=&
 \left[ \int ds \int ds^{\prime} \beta(t-s)
\beta(t^{\prime}-s^{\prime}) \xi(s) \xi(s^{\prime}) \right] P[\xi
(t)] \nonumber \\
&& - \beta(t-t^{\prime}) P[\xi(t)],
\end{eqnarray}

\noindent and
\begin{eqnarray}\label{eq29}
0 &=& \int {\mathcal{D}} \xi \frac{\delta^2 P[\xi(t)]}{\delta
\xi(t)\partial \xi(t^{\prime})},  \nonumber \\
& = & \int ds \int ds^{\prime} \beta(t-s) \beta(t^{\prime}-s^{\prime})
c(s-s^{\prime}) \nonumber \\
&& - \beta(t-t^{\prime}).
\end{eqnarray}

\noindent Eq.(\ref{eq29}) implies
\begin{eqnarray}\label{eq30}
\int ds  \beta(t-s) c(s-s^{\prime}) = \delta (t-s^{\prime}),
\end{eqnarray}

\noindent which shows that the kernel $\beta(s-s^{\prime})$ is the
inverse of the correlation function $c(s-s^{\prime})$. Now, using
Eqs.(\ref{eq27}) and (\ref{eq30}) one may observe that
\begin{eqnarray}\label{eq31}
P[\xi(t)] \xi(t) = -  \int ds^{\prime} c(t-s^{\prime})
\frac{\delta P[\xi(t)]}{\delta \xi(s^{\prime})}.
\end{eqnarray}

The path integral is used to define the probability distribution
functional for $x(t)$, then the solution of Eq.(\ref{eq23})
becomes
\begin{eqnarray}\label{eq32}
P(y,t) = \int {\mathcal D} \xi P[\xi] \delta (y-x(t)).
\end{eqnarray}

\noindent From Eq.(\ref{eq32}) it follows that
\begin{eqnarray}\label{eq33}
\frac{\partial P}{\partial t} = \int {\mathcal D} \xi P[\xi] \left
[ - \frac{\partial}{\partial y}\delta (y-x(t))  {\dot x} \right ],
\end{eqnarray}

\noindent where ${\dot x} = dx/dt$ and can be replaced by
the right hand side of Eq.(\ref{eq23}) to get
\begin{eqnarray}\label{eq34}
\frac{\partial P}{\partial t} &=&  - \frac{\partial}{\partial y}
\int {\mathcal D} \xi P[\xi] \delta (y-x(t)) \left [ W(x)
+\xi(t)\right ], \nonumber \\
&=& - \frac{\partial}{\partial y}[W(y) P] - \frac{\partial}{\partial y}
\int {\mathcal D} \xi P[\xi] \xi(t) \delta (y-x(t)). \nonumber \\
\end{eqnarray}

\noindent Now functional integration by parts yields
\begin{eqnarray}\label{eq35}
&& \int {\mathcal D} \xi P[\xi] \xi(t)\delta (y-x(t))
\nonumber \\
&& =- \int ds^{\prime} c(t-s^{\prime}) \int {\mathcal D} \xi
P[\xi] \frac{\partial}{\partial y} \delta (y-x(t)) \frac{\delta
\xi(t)}{\delta \xi(s^{\prime})}. \nonumber \\
\end{eqnarray}

\noindent Again from Eq. (\ref{eq23}) we have
\begin{eqnarray*}
\frac{d}{dt} \frac{\delta x(t)}{\delta \xi(t^{\prime})} =
W^{\prime}(x)\frac{\delta x(t)}{\delta \xi(t^{\prime})} +
\delta(t-t^{\prime}).
\end{eqnarray*}

\noindent This equation possesses the unique solution
\begin{eqnarray}\label{eq36}
\frac{\delta x(t)}{\delta \xi(t^{\prime})} = \Theta (t-t^{\prime})
\exp \left [\int^t_{t^{\prime}} ds W^{\prime}(x(s)) \right],
\end{eqnarray}

\noindent where $\Theta (t-t^{\prime})$ is defined by
\begin{eqnarray*}
\Theta (t-t^{\prime}) &=& 1, t>t^{\prime} \nonumber \\
                       &=& 1/2,  t=t^{\prime} \nonumber \\
                        &=& 0, t<t^{\prime}.
\end{eqnarray*}

\noindent Now substituting Eq.(\ref{eq36}) into Eq.(\ref{eq35})
and from Eq.(\ref{eq34}) we have
\begin{eqnarray}\label{eq37}
\frac{\partial P}{\partial t} & = & - \frac{\partial}{\partial y}
[W(y)P] + \frac{\partial^2}{\partial y^2} \left \{\int_0^t
ds^{\prime} c(t-s^{\prime}) \right. \nonumber \\
&& \times \int {\mathcal D} \xi P[\xi] \exp
\left [\int^t_{s^{\prime}} ds W^{\prime}(x(s))\right] \nonumber \\
&& \left. \times \delta (y-x(t)) \right \}.
\end{eqnarray}

\noindent Eq.(\ref{eq37}) is not a Fokker-Planck equation. The
second term cannot be reduced to a term containing $P(y,t)$
because of the non-Markovian dependence on $x(s)$ for $s<t$.
Fortunately, in our case $c(t-t^{\prime})=
(\widetilde{D}_R/\tau_R) \exp(-|t-t^{\prime}|/\tau_R)$, where
$\widetilde{D}_R = D_R/\gamma^2$, is an exponentially decaying
function and for large $\gamma$ (as we are dealing with the over
damped case) decays rapidly. We now change the variable
$t^{\prime}= (t -s^{\prime})$ and observe that
\begin{eqnarray} \label{eq38}
&& \int_0^t ds^{\prime} c(t-s^{\prime}) \exp \left
[\int^t_{s^{\prime}} ds W^{\prime}(x(s)) \right] \nonumber \\
&=& \int_0^t
dt^{\prime} c(t^{\prime}) \exp \left[\int_{t-t^{\prime}}^t ds
W^{\prime}(x(s)) \right], \nonumber \\
&\simeq & \frac{{\widetilde D_R}}{\tau_R} \int_0^t
dt^{\prime} \exp (-t^{\prime} / \tau_R) \nonumber \\
&&\times \exp \left [ t^{\prime} W^{\prime}(x(t)) -  \frac{1}{2}
(t^{\prime})^2 W^{\prime \prime}(x(t)) {\dot x}(t) \right].
\end{eqnarray}

\noindent Neglecting the $(t^{\prime})^2$ term which can be shown
to be valid self consistently for small $\tau_R$ under Markov
approximation, \cite{lw,refjw} we get
\begin{eqnarray}\label{eq39}
&& \int_0^t dt^{\prime} c(t^{\prime}) \exp \left
[\int_{t-t^{\prime}}^t ds W^{\prime}(s) \right] \nonumber \\
&\simeq&
\frac{{\widetilde D_R}}{\tau_R} \int_0^t dt^{\prime} \exp \left [
-\frac{t^{\prime}} {\tau_R} + t^{\prime} W^{\prime}(x(t)) \right ],
\nonumber \\
& \simeq & \frac{ {\widetilde D_R} } {1- \tau_R
W^{\prime}(x(t))},
\end{eqnarray}

\noindent for sufficiently large $t$. Substituting Eq.(\ref{eq39})
into Eq.(\ref{eq37}) we obtain the Fokker-Planck-Smoluchowski
equation corresponding to Eq.(\ref{eq19}) as
\begin{eqnarray}\label{eq40}
\frac{\partial P}{\partial t} = \frac{\partial}{\partial x}
\left \{\frac{1}{\gamma} \left [V^{\prime}(x) + \frac{D_R}{\gamma}
\frac{\partial}{\partial x} \left(
\frac{1}{1+(\tau_R/\gamma) V^{\prime\prime}(x)} \right )
\right] P \right\}, \nonumber \\
\end{eqnarray}

\noindent where $P \equiv P(x,t)$, is the probability density of finding the
particle at $x$ at time $t$. Defining an auxiliary function
$G(x)$, Eq.(\ref{eq40}) can be rewritten as
\begin{eqnarray}\label{eq41}
\frac{\partial P}{\partial t} = \frac{\partial}{\partial x}
\left \{\frac{1}{\gamma} \left [V^{\prime}(x) + \frac{D_R}{\gamma}
\frac{\partial}{\partial x} G(x) \right] P \right\},
\end{eqnarray}

\noindent where $G(x)=1/[1+ (\tau_R/\gamma)
V^{\prime\prime}(x) ]$.


\section{Application}\label{application}

\subsection{Solution under periodic boundary condition}

In this subsection we consider the dynamics of a Brownian particle
moving in a periodic potential under a constant external force
$F$. Then the above Fokker-Planck-Smoluchowski equation, Eq.
(\ref{eq41}) reads as
\begin{eqnarray*}
\frac{\partial P}{\partial t} = \frac{\partial}{\partial x}
\left \{\frac{1}{\gamma} \left [V^{\prime}(x) - F +
\frac{D_R}{\gamma} \frac{\partial}{\partial x} G(x) \right] P \right\}.
\end{eqnarray*}

\noindent In the over damped limit the stationary current is given by
\begin{eqnarray}\label{eq42}
J = - \frac{1}{\gamma}\left [V^{\prime}(x) - F +
\frac{D_R}{\gamma} \frac{\partial}{\partial x} G(x) \right]
P_{st}(x),
\end{eqnarray}

\noindent where $P_{st} (x) = P(x,t \rightarrow \infty)$.
Now under symmetric periodic potential with periodicity $2\pi$,
i.e., $V(x) = V(x+2 \pi)$, we may employ the periodic boundary
condition and normalization over one period on $P_{st}(x)$,
\begin{eqnarray}\label{eq43a}
P_{st}(x) = P_{st}(x+2\pi) ,
\end{eqnarray}

\noindent and
\begin{eqnarray}\label{eq43b}
\int_0^{2\pi} P_{st} (x) dx =1.
\end{eqnarray}

\noindent Integrating Eq.(\ref{eq42}) we have the expression of
stationary probability distribution in terms of stationary current
as
\begin{eqnarray}\label{eq44}
P_{st}(x) = \frac{e^{-U(x)} }{G(x)} \left [ G(0)P_{st}(0)
- J \frac{\gamma^2}{D_R} \int_0^x
e^{U(x^{\prime})}dx^{\prime}\right],
\end{eqnarray}

\noindent where
\begin{eqnarray}\label{eq45}
U(x) = \frac{\gamma} {D_R} \int_0^x \frac{V^{\prime}(x^{\prime})
-F}{G(x^{\prime})} dx^{\prime},
\end{eqnarray}

\noindent is the effective potential for the problem. Now applying
the periodic boundary condition Eq. (\ref{eq43a}) we have from
Eq.(\ref{eq42})
\begin{eqnarray}\label{eq46}
G(0) P_{st}(0) = J \frac{\gamma^2}{D_R} \left[ 1 - e^{U(2\pi)}
\right]^{-1} \int_0^{2\pi} e^{U(x)} dx.
\end{eqnarray}

\noindent Using Eq.(\ref{eq42}) and applying the normalization
condition [Eq. (\ref{eq43b})], we get
\begin{eqnarray}\label{eq47}
\int_0^{2\pi} \frac{ e^{-U(x)}}{G(x)} \left [G(0)  P_{st}(0)
 - J \frac{\gamma^2}{D_R} \int_0^x  e^{U(x^{\prime})} dx^{\prime} \right]
 dx = 1. \nonumber \\
\end{eqnarray}

\noindent Elimination of $G(0)  P_{st}(0)$ from Eqs.(\ref{eq46}-\ref{eq47})
gives the expression of stationary current
\begin{eqnarray}\label{eq48}
J = \frac{D_R}{\gamma^2} \left [ \frac{ (1-e^{U(2\pi)})}{
{M}}\right],
\end{eqnarray}

\noindent where
\begin{eqnarray*}
M &=& \left [ \int_0^{2\pi} \frac{e^{-U(x)}}{G(x)} dx \int_0^{2\pi}
e^{U(x^{\prime})}  dx^{\prime}  - \left ( 1-e^{U(2\pi)}\right)
\right. \nonumber \\
&& \left. \times \int_0^{2\pi}  \left \{  \frac{1} {G(x)}e^{-U(x)} \left(
\int_0^{x} e^{U(x^{\prime})}  dx^{\prime} \right) dx \right
\}\right].
\end{eqnarray*}

\noindent Now, the average velocity, $\langle v \rangle = \langle
\dot{x} \rangle$, is given by
\begin{eqnarray*}
\langle v \rangle &=& \frac{1}{\gamma}
\langle (F- V^{\prime}(x) ) \rangle, \\
&=&  \frac{1}{\gamma} \int_0^{2\pi} \left \{ F - V^{\prime}(x)  \right
\} P_{st}(x) dx, \\
&=&  \frac{1}{\gamma} \int_0^{2\pi}
\left \{ \gamma J + \frac{D_R}{\gamma} \frac{\partial}{\partial x}
G(x) P_{st}(x)  \right \} dx,
\end{eqnarray*}

\noindent where we have made use of Eq.(\ref{eq42}). Since
$P_{st}(x)$ and $V(x)$ are periodic functions of $x$, with period
$2\pi$, $\langle v \rangle$ is thus given by the constant
probability current density, multiplied by $2\pi$: $\langle v
\rangle = 2 \pi J$. For the periodic potential
\begin{eqnarray*}
U(2\pi)  = \frac{\gamma}{D_R} \int_0^{2\pi}
\frac{V^{\prime}(x^{\prime}) -F}{G(x^{\prime})} dx^{\prime}
= - F \alpha \text{ (say)}, \\
\end{eqnarray*}

\noindent with
\begin{eqnarray*}
\alpha = \frac{\gamma}{D_R} \int_0^{2\pi}
\frac{dx^{\prime}}{G(x^{\prime})}.
\end{eqnarray*}

\noindent By definition \cite{refjw}, the mobility is given by
\begin{eqnarray*} \mu = \lim_{F\rightarrow 0} \frac{\langle v
\rangle}{F} = 2\pi \; \lim_{F\rightarrow 0} \frac{J}{F}.
\end{eqnarray*}

\noindent If we consider mobility only in linear response regime
\cite{refjw}, the double integral term in the expression of steady
state current becomes
\begin{eqnarray}\label{eq49}
J = \frac{D_R} {\gamma^2} \left [ \frac{ (1-e^{-F\alpha})} {
{M}}\right]
\end{eqnarray}

\noindent where
\begin{eqnarray*}
M &=& \left [ \int_0^{2\pi} \frac{e^{-U(x)}}{G(x)} dx \int_0^{2\pi}
e^{U(x^{\prime})}  dx^{\prime}  - \left ( 1-e^{-F\alpha}\right)
\right. \nonumber \\
&& \left. \times \int_0^{2\pi}  \left \{  \frac{1} {G(x)}e^{-U(x)} \left(
\int_0^{x} e^{U(x^{\prime})}  dx^{\prime} \right) dx \right
\}\right],
\end{eqnarray*}

\noindent vanishes as $F\rightarrow 0$ and hence the mobility is
given by

\begin{eqnarray}\label{eq50}
\mu = \frac{ 2 \pi (D_R/\gamma^2) \alpha} { \int_0^{2\pi}
\frac{ e^{-U(x)}}{G(x)} dx  \int_0^{2\pi} e^{U(x^{\prime})}
dx^{\prime}}.
\end{eqnarray}

\noindent Using Einstein relation \cite{refjw}, the diffusion rate
is given by
\begin{eqnarray}\label{eq51}
{\widetilde D} &=& \mu \left ( k_B T + D \kappa_0^2 \right), \\
\nonumber &=& \frac{4 \pi^2}{\gamma} \left ( k_B T + D \kappa_0^2
\right ) \frac{1} { { \int_0^{2\pi} e^{U(x)} dx \int_0^{2\pi}
\frac{ e^{-U(x)}}{G(x)} dx } }.
\end{eqnarray}

\noindent The above expression for diffusion rate is exact for any
periodic potential and for any Gaussian noise process with
decaying memory kernel. For a simple choice of potential, $V(x)$,
the above expression is analytically tractable. As for example, if
we choose $V(x) = A \cos x$, then to first order in $\tau_R$
(assuming the damping is large)
\begin{eqnarray}\label{eq52}
{\widetilde D} = \frac{k_BT + D \kappa_0^2}{\gamma I_0^2(A)} \left
[ 1 + \tau_R (2\gamma -1) A \frac{I_1(A)}{I_0(A)} \right],
\end{eqnarray}
\noindent where $I_{\nu}$ is the modified Bessel function of order
$\nu$. From Eq.(\ref{eq52}), the diffusion rate is seen to
increase from white to colored noise. The quantity $ A
(I_1(A)/I_0(A))$ is strictly positive, except at $A=0$ where
it is zero. The increase in $\widetilde{D}$ is consistent with
the fact that the diffusion coefficient also increases from white
to colored noise by a factor which is much larger than the
enhancement factor of the potential.

\subsection{Bistable Kinetics}

The dynamics of a Brownian particle in a bistable potential models
several physical phenomena \cite{htb}, and the standard form of
the potential is given by,
\begin{eqnarray}\label{eq53}
V(x) = - \frac{a}{2}x^2 + \frac{b}{4}x^4,
\end{eqnarray}

\noindent which has symmetric minima at $x = \pm  \sqrt{a/b} $
with an intervening local maxima at $x=0$ of relative height $E_b=
a^2/4b$. We consider that our system of interest is moving in a
bistable potential of the form Eq.(\ref{eq53}) and is coupled to a
bath which is modulated by an external Gaussian noise,
$\epsilon(t)$ with the statistical properties stated earlier. The
corresponding Fokker-Planck equation in the over damped limit is
given by Eq.(\ref{eq40}), the explicit form of which with the
potential as in Eq.(\ref{eq53}) reads as
\begin{eqnarray*}
\frac{\partial P}{\partial t} &=& \frac{1}{\gamma} \left [
\frac{\partial}{\partial x}  \left \{ \frac{d}{dx} \left (
-\frac{a}{2}x^2 + \frac{b}{4}x^4 \right)  \right. \right. \nonumber \\
&& \left. \left. + \frac{D_R}{\gamma}
\frac{\partial}{\partial x} \left (  \frac{\gamma}{\gamma + \tau_R
\frac{d^2}{dx^2} \{ -a x^2/2 +bx^4/4 \}  } \right) \right \}
P \right].
\end{eqnarray*}
\noindent At steady state, the above equation reads as
\begin{eqnarray}\label{eq54}
&& \frac{d}{dx} \left ( -ax + bx^3 \right) P(x) \nonumber \\
&&+ \frac{D_R}{\gamma}
\frac{d^2}{d x^2} \left [\frac{\gamma} {\gamma + \tau_R (-a +
3bx^2) } \right]  P (x) = 0.
\end{eqnarray}

\noindent Since at steady state the stationary current vanishes,
Eq.(\ref{eq54}) takes the form
\begin{eqnarray}\label{eq55}
\frac{dP(x)}{dx} + R(x) P(x) =0,
\end{eqnarray}

\noindent where
\begin{eqnarray}\label{eq56}
R(x) = \frac{g^{\prime}(x)}{g(x)} - \frac{h(x)}{D_R g(x)},
\end{eqnarray}

\noindent with
\begin{eqnarray}\label{eq57}
g(x) = \frac{1}{ \gamma - \tau_R (a-3bx^2)},
h(x) = ax -bx^3,
\end{eqnarray}

\noindent and prime ($^{\prime}$) denotes differentiation with respect to
$x$. The solution of Eq.(\ref{eq55}) is given by
\begin{eqnarray}\label{eq58}
P(x) &=& N \left [ |\gamma - \tau_R (a-3bx^2)| \right] \exp
\left [ \frac{a}{2D_R} (\gamma - a \tau_R)x^2 \right.
\nonumber \\
&& \left. + \frac{4ab \tau_R -
b\gamma}{4D_R} x^4 - \frac{\tau_R b^2}{2D_R}x^6  \right],
\end{eqnarray}
\noindent where $N$ is the normalization constant. From
Eq.(\ref{eq58}), we observe that the pre-exponential factor
behaves like a constant in comparison to the exponential factor,
which is an exact statement for $\tau_R =0 $, i.e., when the
external noise $\epsilon(t)$ is $\delta$-correlated.

To this end, following the standard technique \cite{refjx}, the
barrier crossing rate is obtained as
\begin{eqnarray}\label{eq59}
k = \frac{{\sqrt 2} a}{\pi \gamma} \left[ \frac{\gamma - a
\tau_R}{\gamma + 2 a \tau_R}\right] \exp \left [  - \frac{a^2
\gamma}{4b D_R} \left (1+ \frac{2a}{\gamma} \tau_R \right) \right],
\end{eqnarray}

\noindent which is valid in the strong friction regime.


\begin{figure}[t]\label{fig1}
\includegraphics[width=0.6\columnwidth,angle=-90]{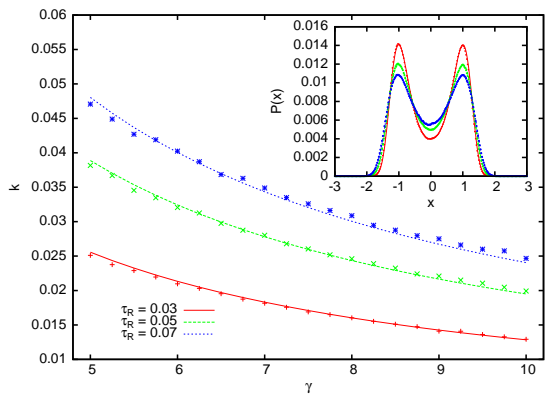}
\caption{(color online) Plot of barrier crossing rate $k$ as a
function of dissipation constant $\gamma$ for various values of
the effective correlation time $\tau_R$. The solid lines are drawn
from the theoretical expression, Eq.(\ref{eq59}) and the symbols
are the results of numerical simulation of Eq.(\ref{eq22}). The
values of the parameters used are $k_BT = D_e = 0.05$, $\tau_e =
0.04 \text{ (red: solid)}, 0.06 \text{ (green: dash)} \text{ and }
0.08 \text{ (blue: dot)}$ and $\kappa_0^2 = 3 \text{ (red:
solid)}, 5 \text{ (green: dash)} \text{ and } 7 \text{ (blue:
dot)}$. Inset: The normalized steady state PDF [ Eq.(\ref{eq58}) ]
using $\gamma = 5.0$ and the same parameter set as in the main
figure.}
\end{figure}

\section{Numerical implementation}\label{secnum}
To check the validity of our analytical result we numerically
simulate the Langevin equation, Eq. (\ref{eq22}) using Heun's
algorithm \cite{heun}. In our simulation, we have always used a
small integration step $\Delta t =0.001$ to ensure numerical
stability. In addition to that all our numerical results have been
averaged over 10, 000 trajectories to obtain a smooth numerical
profile. As mentioned in Sec.II, although we mention the explicit
values of the external noise parameters ($D_e$ and $\tau_e$) used
in the simulation, we interpret our results in terms of the
effective noise parameters $D_R$ and $\tau_R$.

In Fig.(1) we show the profile of escape rate, $k$ as a function
of the dissipation constant, $\gamma$ for different values of the
effective correlation time, $\tau_R$. Numerically, the escape rate
has been defined as the inverse of the mean first passage time
\cite{mfpt}. The values of the different parameters used in the
simulation are given in the figure caption. The profiles show that
the numerical results are in good agreement with the analytical
ones. As expected for a fixed $\tau_R$ value, the escape rate
decreases with $\gamma$ but for a fixed value of $\gamma$, the
escape rate increases with the increase of the effective
correlation time, $\tau_R$. To understand this behavior we
numerically calculate the steady state PDF \cite{mfpt} for a fixed
value of $\gamma$, which is a measure of the dynamics in the
bistable potential. The analytical [ Eq.(\ref{eq58}) ] and
numerical profiles of steady state PDF [see the inset of Fig.(1)]
show that the barrier height decreases as $\tau_R$ increases which
effectively increases the escape rate, $k$. In Fig.(2) we show the
variation of the escape rate with different values of the coupling
term $\kappa_0$ which reflects an increasing trend of $k$ with
$\kappa_0$. The steady state PDF profile accounts for this
behavior with decrease in the barrier height [see the inset].

\begin{figure}[t]\label{fig2}
\includegraphics[width=0.6\columnwidth,angle=-90]{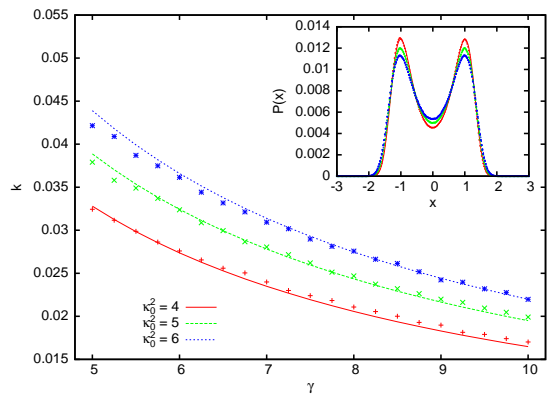}
\caption{(color online) Plot of barrier crossing rate $k$ as a
function of dissipation constant $\gamma$ for various values of
coupling parameter $\kappa_0^2$. The solid lines are drawn from
the theoretical expression, Eq.(\ref{eq59}) and the symbols are
the results of numerical simulation of Eq.(\ref{eq22}). The values
of the parameters used are $k_BT = D_e = 0.05$ and $\tau_e =
0.0625 \text{ (red: solid)}, 0.06 \text{ (green: dash)} \text{ and
} 0.0583 \text{ (blue: dot)}$ so that the effective correlation
time $\tau_R$ always remains 0.05. Inset: The normalized steady
state PDF [ Eq.(\ref{eq58}) ] using $\gamma = 5.0$ and the same
parameter set as in the main figure.}
\end{figure}


\section{Conclusion}\label{conclusion}
A system reservoir model, where the reservoir is modulated
externally by a Gaussian colored noise has been proposed to study
the transport of an overdamped Brownian particle in a periodic
potential. Based on the Fokker-Planck-Smoluchowski description we
calculate the mobility of the Brownian particle in the linear
response regime and using Einstein's relation the diffusion rate
is calculated for any arbitrary periodic potential where the
external driving noise is colored. For a cosine potential we
obtain the diffusion rate in a closed analytical form and observe
that the diffusion rate increases from white to colored noise. As
an immediate application of our formalism, we study the bistable
kinetics of the Brownian particle and demonstrate the dependence
of the correlation time of the external colored noise, by which
the bath is modulated, on the steady state probability density
function and observe the barrier crossing dynamics to obtain the
expression for escape rate. Our analytical result for the escape
rate is then compared with the Langevin simulation result which
shows that both are in very good agreement. To the end it should
be noted that nonlinear system reservoir coupling may also be
considered to obtain a Langevin equation for a Brownian particle
effectively driven by a state dependent colored noise, from which
one may observe various dynamical and kinematical aspects of the
Brownian particle. We hope to address these issues in near the
future.

\begin{acknowledgments}
The authors are grateful to the anonymous referee for his
illuminating and constructive suggestion and information. SC would
like to acknowledge Bengal Engineering and Science University,
Shibpur for financial support (RDO-2/197). Assistance from Dr.
Debi Banerjee for preparing the manuscript is also acknowledged.
\end{acknowledgments}

\end{document}